\documentclass[journal,final,twocolumn]{IEEEtran}
\IEEEoverridecommandlockouts
\usepackage{amsmath}
\usepackage{amssymb}
\usepackage{mathrsfs}
\usepackage{color}
\usepackage{cite}
\usepackage[font=small,labelsep=period]{caption}
\usepackage{graphicx}
\usepackage{subfigure}
\usepackage{algorithmicx}
\usepackage{array}
\usepackage{mdwmath}
\usepackage{mdwtab}
\usepackage{eqparbox}
\usepackage{stfloats}
\usepackage{slashbox}
\usepackage{algorithm}
\usepackage{algpseudocode}
\usepackage{amsmath}

\usepackage{enumitem}

\allowdisplaybreaks
\def\ba{\begin{array}}
\def\ea{\end{array}}
\def\eqa{\begin{eqnarray}}
\def\eqe{\end{eqnarray}}

\def\L2{{\cal L}_2}
\def\L2e{{\cal L}_{2e}}

\def\begequarr{\begin{eqnarray}}
\def\endequarr{\end{eqnarray}}
\def\begequarrs{\begin{eqnarray*}}
\def\endequarrs{\end{eqnarray*}}
\def\begarr{\begin{array}}
\def\endarr{\end{array}}
\def\begequ{\begin{equation}}
\def\endequ{\end{equation}}

\def\begdes{\begin{description}}
\def\enddes{\end{description}}

\def\lef[{\left[\begin{array}}
\def\rig]{\end{array}\right]}

\def\begcen{\begin{center}}
\def\endcen{\end{center}}
\def\endrem{\end{remark}}

\renewcommand{\arraystretch}{1.2}

\begin{document}
\title{Model-Free Load Frequency Control of Nonlinear Power Systems Based on Deep Reinforcement Learning}
\author{
Xiaodi Chen, ~Meng Zhang, ~Zhengguang Wu,
~Ligang Wu, \emph{Fellow, IEEE},
~Xiaohong Guan, \emph{Fellow, IEEE}

\thanks{This work was supported in part by National Natural Science Foundation of China under Grant 62033005, Grant 62273270; in part by Natural Science Foundation of Shaanxi Province under Grant 2023JC-XJ-17, and in part by Natural Science Foundation of Sichuan Province under Grant2022NSFSC0923. This work was supported  by CAAI-Huawei MindSpore Open Fund under Grant CAAIXSJLJJ-2022-001A. (Corresponding author: Meng Zhang.) }
\thanks{X. Chen, M. Zhang, and X. Guan are with the  School of Cyber Science and Engineering, Xi'an Jiaotong University, Xi'an 710049, China.
(e-mail: cxd0524@stu.xjtu.edu.cn; mengzhang2009@xjtu.edu.cn; xhguan@xjtu.edu.cn).}
\thanks{Z. Wu is with the College of Control Science and Engineering, Zhejiang University, Hangzhou Zhejiang, 310027, China (e-mail: nashwzhg@zju.edu.cn).}
\thanks{L. Wu is with the Department of Control Science and Engineering, School
of Astronautics, Harbin Institute of Technology, Harbin 150001, China. (e-mail: ligangwu@hit.edu.cn).}
\markboth{IEEE}%
{Shell \MakeLowercase{\textit{et al.}}: Bare Demo of IEEEtran.cls for Journals}
}
\maketitle
\begin{abstract}

Load frequency control (LFC) is widely employed in power systems to stabilize frequency fluctuation and guarantee power quality.
However, most existing LFC methods rely on accurate power system modeling and usually ignore the nonlinear characteristics of the system, limiting controllers' performance.
To solve these problems, this paper proposes a model-free LFC method for nonlinear power systems based on deep deterministic policy gradient (DDPG) framework. The proposed method establishes an emulator network to emulate power system dynamics. After defining the action-value function, the emulator network is applied for control actions evaluation instead of the critic network. Then the actor network controller is effectively optimized by estimating the policy gradient based on zeroth-order optimization (ZOO) and backpropagation algorithm.
Simulation results and corresponding comparisons demonstrate the designed controller can generate appropriate control actions and has strong adaptability for nonlinear power systems.
\end{abstract}


\IEEEpeerreviewmaketitle
\section{Introduction}\label{sec1}

{
Frequency serves as a vital indicator of the quality of electrical energy, reflecting the balance between power demand and generation. Any mismatches between power demand and generation can result in frequency deviations, which pose a risk to the stability and reliability of the entire power system. Load frequency control (LFC) emerges as a linchpin in control of power systems, aiming to minimize frequency variations.} By continuously and precisely adjusting the power generation to match the load demand, LFC maintains the system frequency within a specific nominal range, thereby ensuring the stable operation of the power system \cite{9205594,8100916,9626584}.


Numerous results on LFC have been reported in recent decades,  for instance,
\cite{cai2016new}
{presented a detailed analysis on the energy and structure of LFC and successfully achieved the decoupling of tie-line power flow by devising a PID controller based on the Port-Hamiltonian system and cascade system.
\cite{chuang2016robust}
proposed a robust controller based on a minimum order dynamic LFC model and minimized frequency deviations significantly by utilizing a sector-bounded $H\infty$ control approach.
\cite{su2017event}
developed a linear mathematical power system model and designed a discontinuous controller based on sliding-mode control to ensure system stability and robustness.
\cite{al2021design}
developed a single-input fuzzy logic controller using the signed distance method and successfully regulated the system frequency to its desired value under various power system states.
\cite{siti2022application}
incorporated an energy storage system into the control loop and applied model predictive control to optimize frequency deviations by balancing the control behavior of power systems.
\cite{zhang2022reliable, wu2019adaptive, zhang2022distributed,liu2022hybrid} applied event-triggered control to the LFC, resulting in significant reductions in real-time system measurement communication costs.
Although these aforementioned control methods have made remarkable achievements on LFC, they are generally designed based on physical system models and require high real-time computational sources.}
In addition, increasing penetration of renewable energy brings more uncertainties for power system modeling. Therefore, conventional LFC methods heavily relying on accurate system models are difficult to achieve desired control effects for power systems with high uncertainties.

Fortunately, reinforcement learning including deep reinforcement learning (DRL) approaches \cite{zhang2022attack} have emerged as promising solutions to deal with uncertainties of power systems due to their powerful searching and learning ability. To mention a few,
\cite{singh2017distributed} introduced a multi-agent reinforcement learning framework consisting of an estimator agent and a controller agent, where the controller parameters were optimized based on particle swarm optimization to enhance the dynamic performance of the control system. \cite{yin2017artificial} introduced an artificial emotional reinforcement learning controller combining a mechanical logical part and a humanistic emotional part to generate control signals for different scenarios.
\cite{yan2018data} utilized stacked denoising auto-encoders to extract features and improved system performance by deriving control actions in a continuous domain.
\cite{xi2020multi} designed an automatic power generation control algorithm based on double Q-Learning, mitigating the over-estimation of action value and the phenomenon of over-fitting.
\cite{rozada2020load}
modeled each generation unit control agent as a recurrent neural network to enhance the long-term performance of the controller.
\cite{li2022coordinated}
proposed an evolutionary imitation curriculum DRL algorithm that combined imitation learning and curriculum learning to derive optimal coordinated control strategies adaptively.
These methods in the aforementioned literature perform well for linearized LFC models.
However,  most of the results neglect nonlinear behaviors exhibited by the generators and some still rely on power system dynamic equations, which motivate us to carry out this paper.

In response to above discussions, this paper proposes a model-free LFC method based on DDPG for nonlinear power systems. The proposed method builds an emulator network with a historical database {instead of critic network}  to obtain the policy gradient based on zeroth-order optimization and backpropagation algorithm.
With the calculated gradient, the DDPG agent converges to the optimal control policy, where frequency deviations are minimized.
The main contributions of the paper are summarized as follows:
\begin{itemize}
\item
A model-free nonlinear LFC method is proposed to utilize the emulator network to emulate the dynamic characteristics of the power system, thereby avoiding the need for precise modeling of the system.
Validation results show that the emulator network has a remarkable ability to extract the features of nonlinear generator behaviors and performs well in scenarios where the external environment is challenging to be modeled precisely.
\item
The design of the policy gradient with the well-defined action-value function grants the calculated policy gradient practical physical significance. The policy gradient in this paper offers a more explicit updating direction to the actor network compared to the policy gradient obtained by the critic network, which significantly improves the effectiveness of reinforcement learning training.
\item
By incorporating zeroth-order optimization into the calculation of the policy gradient, the issues of gradient vanishing and exploding commonly encountered in deep neural networks are effectively mitigated. The zeroth-order gradient is calculated to replace the policy gradient obtained through backpropagation in deep networks, ensuring stable and reliable parameter updates of the agent during training.

\end{itemize}

The remaining sections of the paper are organized as follows. Section \ref{sec2} introduces the linearized LFC model and the nonlinear generator models. In Section \ref{sec3}, the framework of the proposed method and the corresponding algorithm are presented in detail.
Furthermore, in Section \ref{sec4}, the effectiveness and advantages of the proposed method are tested on both linearized and nonlinear LFC models.
Finally, the conclusion is presented in Section \ref{sec5}.

\section{Power system frequency response model}\label{sec2}
This section provides a brief introduction to both the linearized LFC model and the nonlinear LFC model.
It outlines key characteristics and components of each model, setting the foundation for subsequent discussions and analysis.
\begin{figure}[ht]
  \centering\includegraphics[width=8.89cm]{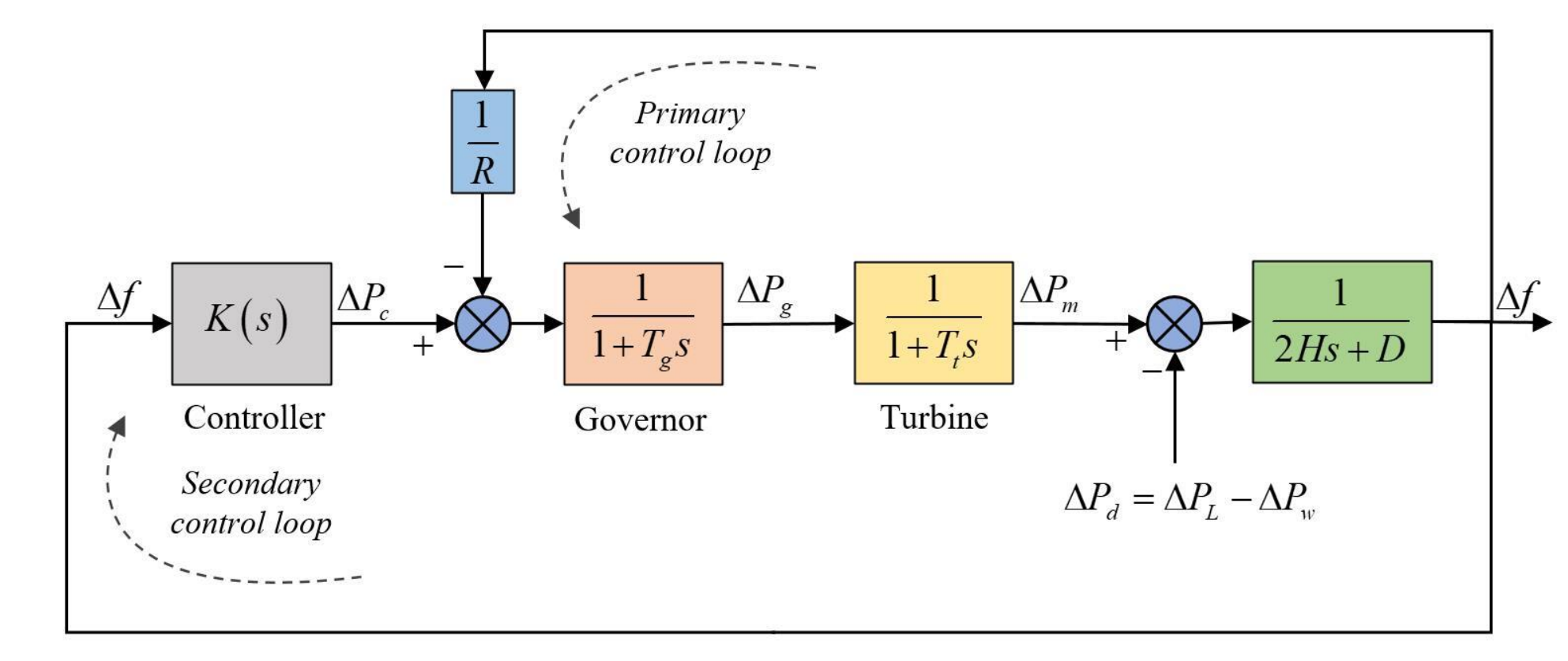}
  \caption{Schematic diagram of the linearized LFC model.}
  \label{fig1}
\end{figure}
\subsection{Linearized LFC Model}
This paper treats stochastic renewable energy as a random load in the LFC model. Fig. \ref{fig1} shows the typical linearized LFC model where wind power injection $\Delta P_{w}$ is considered. According to the schematic diagram of Fig. \ref{fig1}, the LFC system comprises a generator, governor, turbine, and other links in the secondary control loop\cite{bevrani2014robust}. The system dynamics can be expressed as
\begin{equation}
    \label{Eq1}
    \Delta \dot{f} = \frac{1}{2H}\left ( \Delta P_m  - \Delta P_d \right ) - \frac{D}{2H}\Delta f
\end{equation}
\vspace{0.98mm}
\begin{equation}
    \label{Eq2}
    \Delta \dot{P_m} = \frac{1}{T_t} \Delta P_g - \frac{1}{T_t}\Delta P_m
\end{equation}
\vspace{0.98mm}
\begin{equation}
    \label{Eq3}
    \Delta \dot{P_g} = \frac{1}{T_g} \Delta P_c - \frac{1}{RT_g}\Delta f - \frac{1}{T_g} \Delta P_g,
\end{equation}
where $\Delta f$, $\Delta P_m$,  and $\Delta P_g$ represent the deviation of frequency, generator mechanical output, and valve position, respectively. $\Delta P_d$ represents random load fluctuation which includes both load change $\Delta P_L$ and wind power injection $\Delta P_{w}$. $T_t$ represents the time constant of the turbine. $T_g$ represents the time constant of the governor. $H$ represents the inertia constant. $D$ represents the generator damping coefficient. $\Delta P_c$ represents the generator command, i.e.,  the {control action} calculated by the feedback signal $\Delta f$.

\subsection{Nonlinear LFC Model}
\begin{figure}[ht]
  \centering\includegraphics[width=8.89cm]{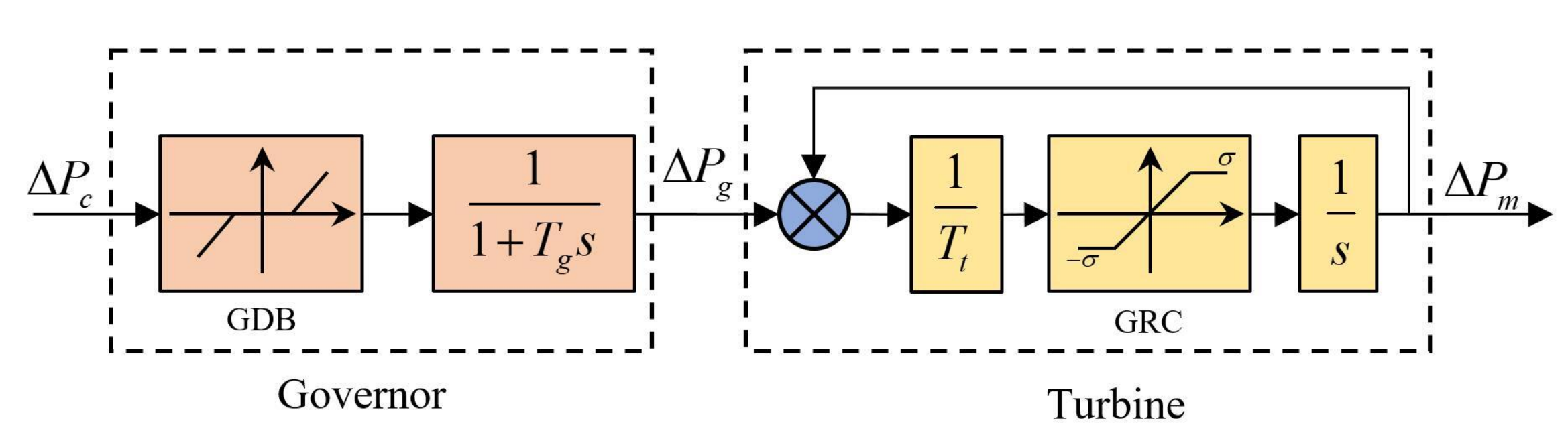}\\
  \caption{Nonlinear behaviors of the generator: GDB and GRC.}
  \label{fig2}
\end{figure}

For the practical application of the proposed method, it is crucial to accurately model the nonlinear characteristics imposed by the physical dynamics of power systems. This paper focuses on two fundamental nonlinear constraints of the generator: the governor dead band (GDB) and the generation rate constraint (GRC). These constraints are illustrated in Fig \ref{fig2} \cite{yan2020multi}, and their inclusion in the model is essential for capturing the realistic behavior of the power system.

GDB is a common nonlinear behavior of generators in power systems, and the effect of  GDB on valve position $P_g(t)$ is shown in Eq. (\ref{EqGDB}){\cite{yang2022data}}. Exaggerated responses of the valve can be avoided by manually setting the GDB, which is significant for maintaining system frequency stability against load disturbance. When the input signal of the speed governor changes, it may not correspondingly act until the input signal is above a specific value. That is, the valve position takes no immediate changes in a magnitude of a speed change defined as dead zone\cite{bevrani2014robust}.
\begin{equation}
    \label{EqGDB}
    \begin{split}
     P_g(t) = & \max \left ( 0, \Delta P_c(t) - \kappa \right )\\
    & + \min(0, \Delta P_c(t) + \kappa),
    \end{split}
\end{equation}
where $\max(x,y)$ means the maximum value between $x$ and $y$  and $\min(x,y)$ means the minimum value. {$\kappa$ represents the dead band of the governor.}

GRC is another typical nonlinear constraint of generators in power systems, and the effect of  GRC on generator mechanical output $P_m(t)$ is shown in {Eq. (\ref{EqGRC})} {\cite{yang2022data}}.
\begin{equation}
    \label{EqGRC}
   -\sigma \leqslant \frac{\mathrm{d}P_m}{\mathrm{d}t}\leqslant \sigma,
\end{equation}
where $\sigma$ represents the threshold value.
 When a large load disturbance occurs, the generator needs to rapidly produce the corresponding power generation so that the power generation and load achieve a new dynamic balance. However, due to the limitation of thermal motion, mechanical structure, and other factors, there is a threshold value that limits the variation of the generation rate and makes the generator difficult to follow the fluctuations of load. That is, GRC hampers the regulation of generation power.



\section{Main results}\label{sec3}
In this section, the main results of this paper, i.e.,  the model-free DDPG method aimed to deal with the nonlinear power system LFC problem {are} presented.

\subsection{Model-free DDPG method for LFC}
The DDPG method is developed within the actor-critic (AC) framework, which consists of actor and critic networks. The actor network is responsible for generating actions and interacting with the environment. Meanwhile, the critic network evaluates the performance of the actor network and provides guidance for the actor's actions in the next stage based on the action-value function. This interaction between the actor and critic networks enables the DDPG method to learn and optimize the control policy.

\begin{figure}[ht]
  \centering\includegraphics[width=8.5cm]{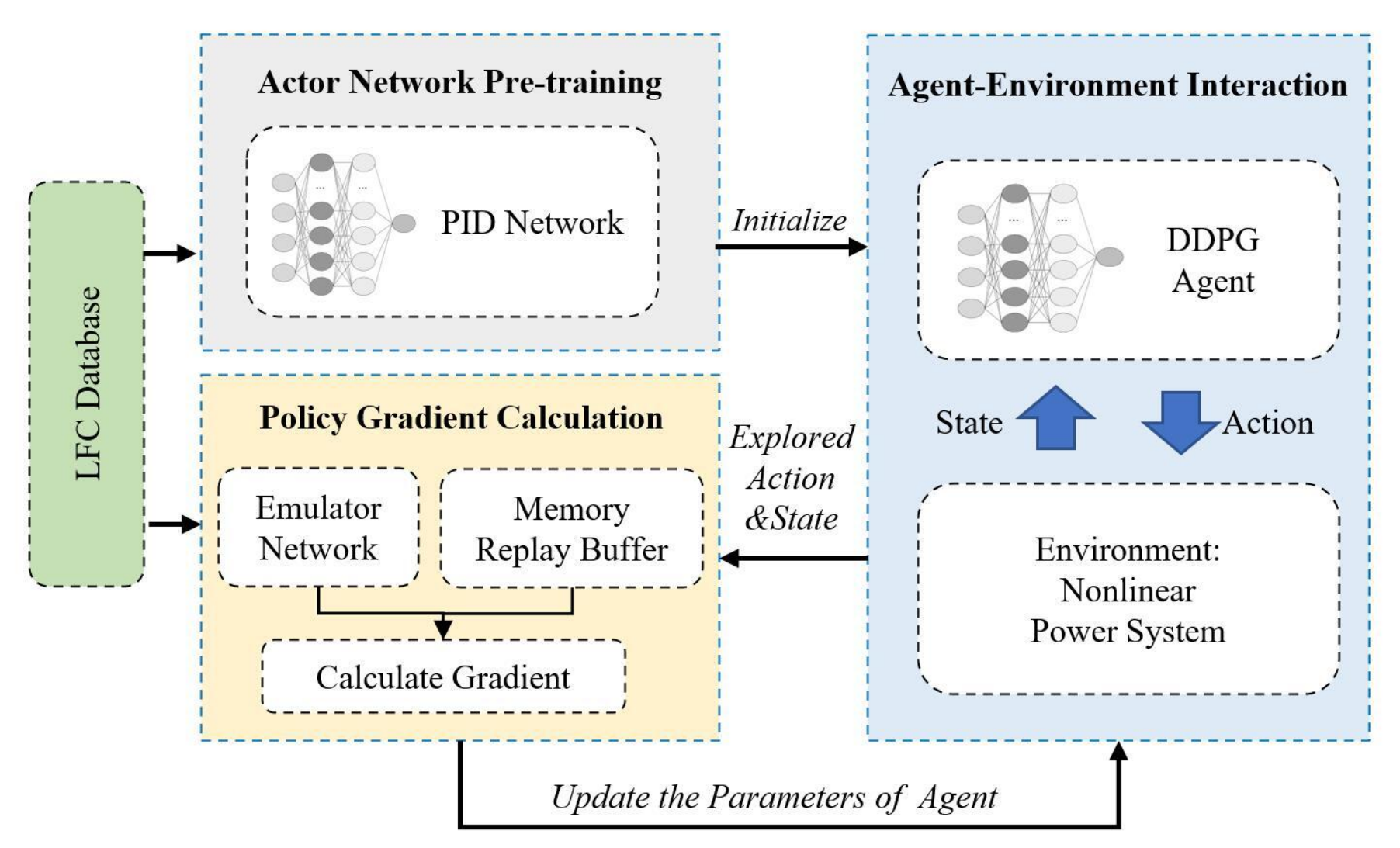}\\
  \caption{Framework of model-free DDPG method.}
  \label{fig3}
\end{figure}

Fig. \ref{fig3} illustrates the schematic of the proposed model-free DDPG method.
The policy agent (i.e., actor network) developed by deep neural network (DNN) is built as the controller. The actor network generates the generation command (i.e., $\Delta P_c$) given the state of the environment to keep the frequency constant. The following equation defines the observation of the agent.
\begin{equation}
    \label{Eqstate}
    {o(s) = [\Delta f(s),\Delta f(s)/s,s\Delta f(s)]},
\end{equation}
where $\Delta f(s)$, $\Delta f(s)/s$, $s\Delta f(s)$ are the proportion, integrate and derivate of frequency deviation $\Delta f$, respectively. It is noted that the $s$ in the bracket of Eq. (\ref{Eqstate}) and Fig. \ref{fig1} represents s-domain after Laplace transformation, which is different with $s$ or $s_{i}$ throughout the paper. {After the input layer of the actor network, a Multi-Layer Perceptron (MLP) is employed. The MLP consists of two fully connected layers, each with 256 neurons.}

Meanwhile, the critic network is replaced by another DNN, named emulator network in this paper. {The emulator network maps the relationship between control actions and system state based on supervised learning.} The nonlinear power system LFC model defined by Eq. (\ref{Eq1})-(\ref{EqGRC}) is set as the external environment of the DRL agent.
During the model-free DDPG learning stage, the agent (i.e., controller) makes action exploration to interact with the environment at each learning iteration and stores transitions into the memory reply buffer.
With the emulator network finishing training, the policy gradient is obtained based on the zeroth-order gradient for the agent updates.

\subsection{Design of Policy Gradient}
Considering the optimization objective of the LFC,
we define the following action-value function $Q_\mu$ to evaluate the control command generated by the policy agent based on the state of the power system.
\begin{equation}
    \label{Eq5}
    Q_\mu \left ( s,a\right) = - \Delta f_{t+1}^2(s,a),
\end{equation}
where $\Delta f_{t+1}$ is the frequency deviation at time step $(t+1)$, $s$ is the state of the power system, $a$ is the control action and
$\mu$ represents the current control policy of actor network.

The objective of updating the actor network's parameter which is denoted as $J$, is to optimize the expectations of the action value $Q_\mu$  shown as
\begin{equation}
    \label{otherEq_maxE}
    J(\theta_\mu) =  E\left [-\Delta f^2\mid s_i,a_i;\theta_\mu \right],
\end{equation}
where $\theta_\mu$ including weighting matrix and biased vector $\theta_\mu = [W^T, b]$ represents the parameter of actor network {and $E\left [ \cdot\right]$ represents the expectation.}
During each iteration of exploration, the agent generates one transition for its learning process. In order to minimize the deviation of frequency, the parameters of the agent are updated for all transitions as
\begin{equation}
    \label{otherEq_maxE2}
    \mathop{\max}_{\theta_\mu}J(\theta_\mu).
\end{equation}

For actor network updating, the experience relay (ER) technique is employed in which a mini-batch dataset is randomly selected from
the reply buffer.  The following policy gradient is calculated based on the chain rule to maximize expectations of $Q_\mu$, representing the optimization objective of the agent \cite{lillicrap2015continuous}.
\begin{equation}
    \label{Eq4}
    \nabla _{\theta _\mu} J \approx \frac{1}{M} \sum_{i=1}^{M}\left [\nabla _a Q_\mu \left ( s, a\right) \mid _{s=s_i, a=\mu (s_i\mid\theta_\mu)}\nabla_{\theta_\mu }\mu\left ( s \mid \theta_\mu \right )\right ],
\end{equation}
where $\nabla _{\theta_\mu}J$ is the sampled policy gradient for the policy agent parameter updating. $M$ is the mini-batch sampling size, and $\mu\left ( s\mid \theta_\mu \right )$ represents the policy to generate the control actions based on the current parameter of the actor network.



\begin{figure}[b]
  \centering\includegraphics[width=8.9cm]{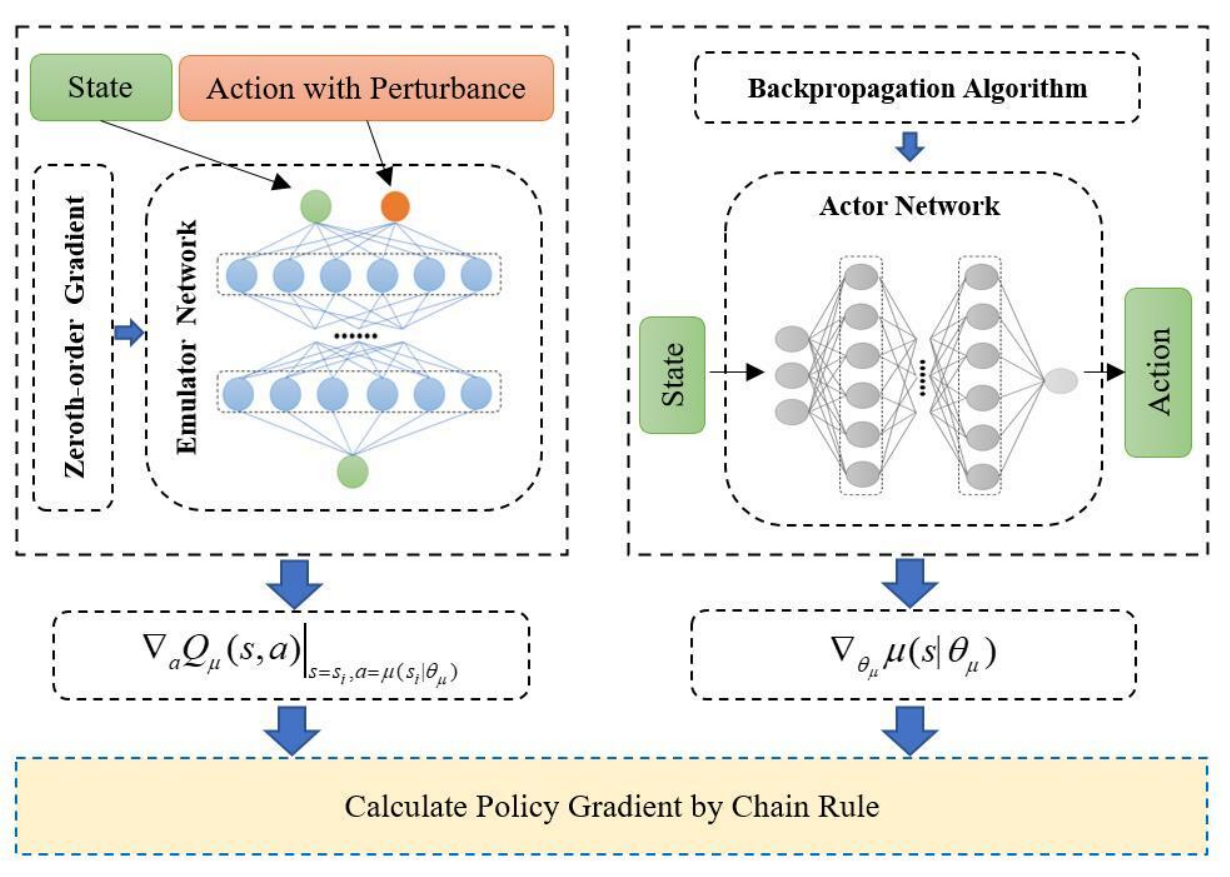}\\
  \caption{{The calculation process of policy gradient.}}
  \label{figprogress}
\end{figure}

\subsection{Calculation of Policy Gradient}\label{subsectionGradient}
After exploring and storing the transitions into the memory replay buffer, the DDPG agent updates {parameters $\theta$} with the policy gradient that determines the training result of the agent. In previous works, the policy gradient is usually obtained by training critic network \cite{lillicrap2015continuous} or derived by model formulas \cite{yan2018data},
where either of them has its own limitations. Training the critic network can be time-consuming and the network may converge to sub-optimal solutions or even oscillate. Moreover, the estimated policy gradient from critic networks may have high variance, which can affect the stability of the learning process.
On the other hand, deriving the policy gradient using model formulas heavily relies on the accuracy of the system model and its equivalent parameters, which makes it unsuitable for LFC in power systems with parameter uncertainties and nonlinear characteristics since the model accuracy is difficult to be guaranteed. Additionally, the derivation process of the action-value function $Q_\mu$ can be complex or difficult to accomplish, further limiting the practicality of this method.
To address these limitations, the proposed model-free DDPG method designs  {an} emulator network to obtain the policy gradient.

{Fig. \ref{figprogress} shows the calculation process of the policy gradient in Eq. (\ref{Eq4}), where two components, i.e., $\nabla _a Q_\mu \left ( s, a\right) \mid _{s=s_i, a=\mu (s_i\mid\theta_\mu)}$ and $\nabla_{\theta_\mu }\mu\left ( s \mid \theta_\mu \right )$ need to be calculated. Firstly, }we take the gradient of Eq.(\ref{Eq5})  to determine the gradient of the action-value function with respect to the control action, which can be expressed as
\begin{equation}
    \label{Eq6}
    \frac{\partial Q_\mu}{\partial a_t}= -2\Delta f_{t+1}\frac{\partial \Delta f_{t+1}}{\partial a_t}.
\end{equation}

To estimate the gradient of the frequency deviation $\Delta f_{t+1}$ with respect to the control action $a_t$, an emulator network is established to serve as an approximation for mapping the state-action pairs to frequency deviations, enabling the estimation of the system dynamics. It provides a means to approximate the relationship between the actions taken by the agent and the resulting frequency deviation in the next state. {The emulator network takes both the state and action as inputs. To adjust the dimensionality of the output vector, an MLP is utilized. This MLP is structured with two fully connected layers, each featuring 256 neurons. The final output of the emulator network is the frequency deviation.} That is
\begin{equation}
    \label{Eqvarphi}
    \Delta \hat{f}_{t+1} = \varphi \left ( s_t,a_t\right ),
\end{equation}
where $\Delta \hat{f}_{t+1}$ is the estimated value of the emulator network and $\varphi (\cdot)$ is the nonlinear mapping from  $a_t$ and $s_t$ to $\Delta \hat{f}_{t+1}$.
A training set is sampled from an LFC database, where the input signal defined by Eq. (\ref{Eqstate}) and the output signal (i.e., control action) of a tuned PID controller are calculated and stored, to finish the supervised training of the emulator network.
By extracting features from $s_t$ and $a_t$, the emulator network provides estimated value $\Delta \hat{f}_{t+1}$. With the emulator network replacing the critic network, the action-value function shown as Eq. (\ref{Eq5}) is utilized to evaluate actions and compute the first component $\nabla_a Q_\mu \left ( s, a\right) \mid_{s=s_i, a=\mu (s_i\mid\theta\mu)}$ in Fig. \ref{figprogress}.
{
However, as a type of DNN, the emulator network faces the challenge of gradient vanishing and gradient explosion during the gradient backpropagation process, primarily due to multiple consecutive layers of the network. Specifically, the policy gradient obtained through the emulator network based on backpropagation could potentially lead the actor network fall into states of gradient vanishing and gradient explosion more easily, which may have a detrimental impact on the final training result of the agent.}

{To mitigate the challenge of gradient vanishing and gradient explosion, the zeroth-order optimization approach \cite{chen2017zoo}  is employed to approximate the policy gradient. Zeroth-order optimization is an essential branch of mathematical optimization that relies on the parameter sampling technique to determine the direction of parameter updates. In this paper, we utilize a limit-based approach (i.e., the definition of gradients) and directly perform random sampling in the parameter space to obtain estimations of the gradient.
For the emulator network, the zeroth-order gradient with respect to $a_t$ at point $(s_t, a_t)$ is defined as
\begin{equation}
    \label{otherEq_zoo}
    \frac{\partial \Delta f_{t+1}}{\partial a_t} \approx E_{u\sim p(u)}\left [ \frac{\varphi(s_t,a_t+\epsilon u)-\varphi(s_t,a_t-\epsilon u))}{2\epsilon}u \right ],
\end{equation}
where $\epsilon$ is a given small positive value. $p(u)$ is a distribution whose mean is 0 and covariance is an identity matrix, typically represented by a standard normal distribution. $E_{u\sim p(u)}$ is the expectation of symmetric difference quotient\cite{lax2014calculus} in the equation.}

By sampling multiple points from the distribution $p(u)$ and introducing a small disturbance $\epsilon u$ to the trained emulator network as shown in Eq. (\ref{otherEq_zoo}),  we can calculate $\nabla_a Q_\mu \left ( s, a\right) \mid_{s=s_i, a=\mu (s_i\mid\theta\mu)}$ by substituting Eq. (\ref{otherEq_zoo}) into Eq.(\ref{Eq6}).
For the other component $\nabla_{\theta_\mu }\mu\left ( s \mid \theta_\mu \right )$ in Fig.\ref{figprogress}, we utilize the backpropagation algorithm to determine the gradient of actions with respect to the parameters of the actor network.
By calculating the policy gradient as above procedures, the agent's parameters are updated iteratively using Eq. (\ref{Eqtheta}). 
\begin{equation}
    \label{Eqtheta}
    \theta ^{(k+1)}_\mu=  \theta_\mu ^{(k)}+\eta \nabla _{\theta_\mu ^{(k)}}J,
\end{equation}
where $\theta_\mu^{(k)}$ and $\theta\mu ^{(k+1)}$ are the parameters of actor network in $k$th and $(k+1)$th learning iteration, respectively. $\eta$ represents the learning rate and $\tau$ is the parameter of soft update. The control performance can be optimized with agent parameters updated as Eq. (\ref{Eq5})-(\ref{Eqtheta}).

\subsection{Algorithm}
The algorithm of the proposed model-free DDPG LFC method for nonlinear power systems can be described as Algorithm 1.
{
It is important to highlight that random initialization of the actor network can significantly increase the training duration of the agent, due to the possibility of generating control actions that deviate from the normal action space resulting in abnormal system states. The corresponding abnormal experiences can produce extra complexity for the emulator network to evaluate the control actions, leading to unnecessary oscillations during the actor network training. To tackle this issue and expedite the training process, an effective solution is to pre-train the actor network.
By starting with a PID-based actor network, the DDPG algorithm benefits from a smoother and more informed exploration of the control space.
This involves sampling a training set from the LFC database established in subsection \ref{subsectionGradient} based on a tuned PID controller and using it to build a DNN named the PID network. The parameters of the actor network are then initialized using the pre-trained PID network, which helps to provide a more reasonable starting point for the actor network and facilitates faster convergence during the training
process.}

Subsequently, the agent interacts with the environment using the initial control policy status, and the transitions $(s_t, a_t, s_{t+1}, r_t)$ are stored in the memory replay buffer. This replay buffer serves as a reservoir of experiences that can be used for training.
Once the exploration phase is complete, a random mini-batch of $M$ transitions is sampled from the buffer. The actions in the sampled transitions are then evaluated using the trained emulator network. By employing zeroth-order optimization, the symmetric difference is computed. This calculation helps to estimate the policy gradient, which is essential for updating the agent's parameters.
The agent repeats the exploration process and parameter update, continuously striving to improve its control policy. This iterative process allows the agent to progressively update its parameters and move towards a better control policy status. As the parameters are sufficiently updated over multiple iterations, the DDPG agent gradually converges to an optimal status where it can effectively regulate the frequency of the power system.

\begin{algorithm}
  \label{a1}
  \caption{Model-free DDPG training process for LFC}
  \begin{algorithmic}[1]
    \Require
      The benchmark LFC system;
      Trained emulator network and PID network;
      Learning rate $\eta$;
    \Ensure
      Parameters of actor network $\theta^\mu$;
    \State  Initialize actor network $\mu(s|\theta^\mu)$ with parameters of trained PID network;\\
    Initialize memory reply buffer    {with the LFC database};
    \For{episode $k=1:P$}
        \State  {Apply an Ornstein-Uhlenbeck process to generate noise $\mathcal{N}$ for the exploration of the agent\cite{lillicrap2015continuous};}\\
        \quad \ Observe initial state $s_i$;
        \For{time $t=1:T$}\\
            \qquad \quad Generate control action $a_t=\mu(s_t|\theta_t^\mu)+ \mathcal{N}$;\\
            \qquad \quad Interact with the environment through  $a_t$;\\
            \qquad \quad  {Observe new state $s_{t+1}$}; \\
            \qquad \quad Store transition $(s_i, a_i, s_{i+1}, r_i)$ in reply buffer;
        \EndFor\\
        \quad \ Sample a random mini-batch of $M$ transitions $(s_i, a_i, s_{i+1}, r)$ from reply buffer;\\

        \quad \ Calculate $\nabla_a Q^\mu\left ( s_t,a_t\right )$ by the trained emulator network as
        {
        \begin{equation}
            \nonumber
                \frac{\partial Q_\mu}{\partial a_t}= -2\Delta f_{t+1}\frac{\partial \Delta f_{t+1}}{\partial a_t}
        \end{equation}
        \begin{equation}
            \nonumber
            \frac{\partial \Delta f_{t+1}}{\partial a_t}\approx E_{u\sim p(u)}\left [ \frac{\varphi(s_t,a_t+\epsilon u)-\varphi(s_t,a_t-\epsilon u))}{2\epsilon}u \right ]
        \end{equation}}\\
        \quad \ Calculate $\nabla _{\theta^{(k)}} \mu\left ( s_i\mid \theta^{(k)} \right )$ based on backpropagation algorithm;\\
        \quad \ Calculate $\nabla _{\theta^\mu}J$ by the chain rule as
        {
        \begin{equation}
            \nonumber
    \nabla _{\theta _\mu} J \approx \frac{1}{M} \sum_{i=1}^{M}\left [\nabla _a Q_\mu \left ( s, a\right) \mid _{s=s_i, a=\mu (s_i)}\nabla_{\theta_\mu }\mu\left ( s \mid \theta_\mu \right )\right ]
        \end{equation}
        }\\
        \quad \ Update the actor network
         \begin{equation}
        \nonumber
            \begin{split}
                \theta_\mu \leftarrow \theta _\mu + \eta \cdot \nabla _{\theta_\mu} J
            \end{split}
        \end{equation}
    \EndFor
  \end{algorithmic}
\end{algorithm}

\section{Case study}\label{sec4}

\begin{table}[b]
  {
 \caption{Parameters of the power system.}
 \label{tab1}
 \renewcommand\arraystretch{2.5}
 \begin{tabular}{c|c c c c c}
   \hline
   \hline
   Parameter & $T_g$(s) & $T_t$(s) & $H$(p.u./Hz) & $D$(p.u./Hz) & $R$(Hz/p.u.)  \\
   \hline
   Value & 0.10 & 0.40 & 0.0833 & 0.0015 & 0.33 \\
   \hline
   \hline
 \end{tabular}
 }
  \end{table}
The proposed method is tested on a power system whose parameters are given in Table \ref{tab1} \cite{yan2020multi} with the Simulink toolbox of MATLAB. {
We utilize the Simulink toolbox to construct integral and derivative modules, enabling us to obtain the integration and derivation of the frequency deviation during the interaction with the environment.}
{For the controller updating in simulations, we set the number of episodes, the size of the memory replay buffer  and the learning rate $\eta$ as  100, 8000, and 0.0005, respectively. The disturbance value $\epsilon$ is set as 0.0001, whose specific value is chosen to ensure a numerical perturbation of approximately 3\% in the inputs of the emulator network. The choice of approximately  3\% is made based on the characteristics of the LFC model parameters we employed, where the perturbation level strikes a balance between introducing sufficient disturbance for effective learning and maintaining the stability of the training process.} An optimized PID controller in \cite{daneshfar2012multiobjective} and a state-of-art model-based DDPG approach in \cite{yan2018data} are compared with the proposed method. And a PID controller, whose data builds the LFC database, is also used for {comparison}, and its gains are tuned manually. The memory replay buffer is initialized by the LFC database.
{The dynamic responses of frequency deviation serve as a crucial metric for assessing the overall performance of the power system.}
To intuitively demonstrate the control effectiveness of the DDPG agent, the cumulative reward $\mathcal{R}$ of the agent during a training iteration is defined as
\begin{equation}
    \label{otherEq_reward}
    \mathcal{R} = -\sum_{t=1}^{T}\left | \Delta f_t\right |,
\end{equation}
where $T$ is the total time of LFC simulation.

\subsection{Results on  linearized LFC Model}
To show the proposed model-free DDPG-based controller also performs well for linearized power systems, the performance comparison between some existing controllers and the proposed controller on the linearized power system model is presented in this subsection.
\begin{figure}[ht]
  \centering\includegraphics[width=9.5cm]{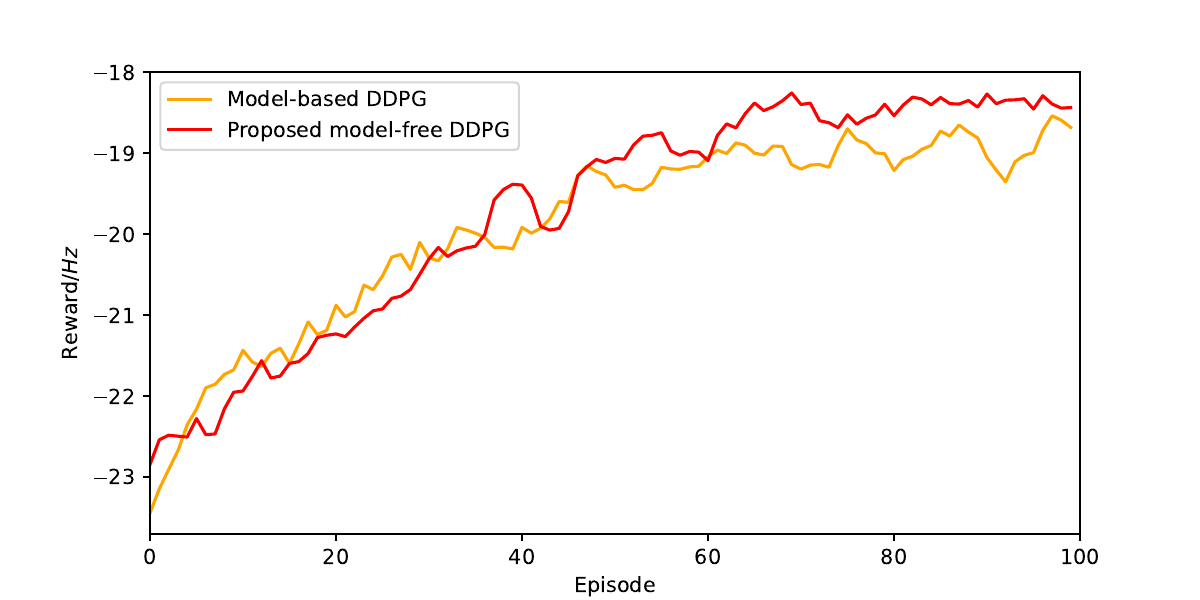}\\
  \caption{{The rewards during training process on linearized power system.}}
  \label{otherfigure_linearreward}
\end{figure}

\begin{figure}[ht]
  \centering\includegraphics[width=9.5cm]{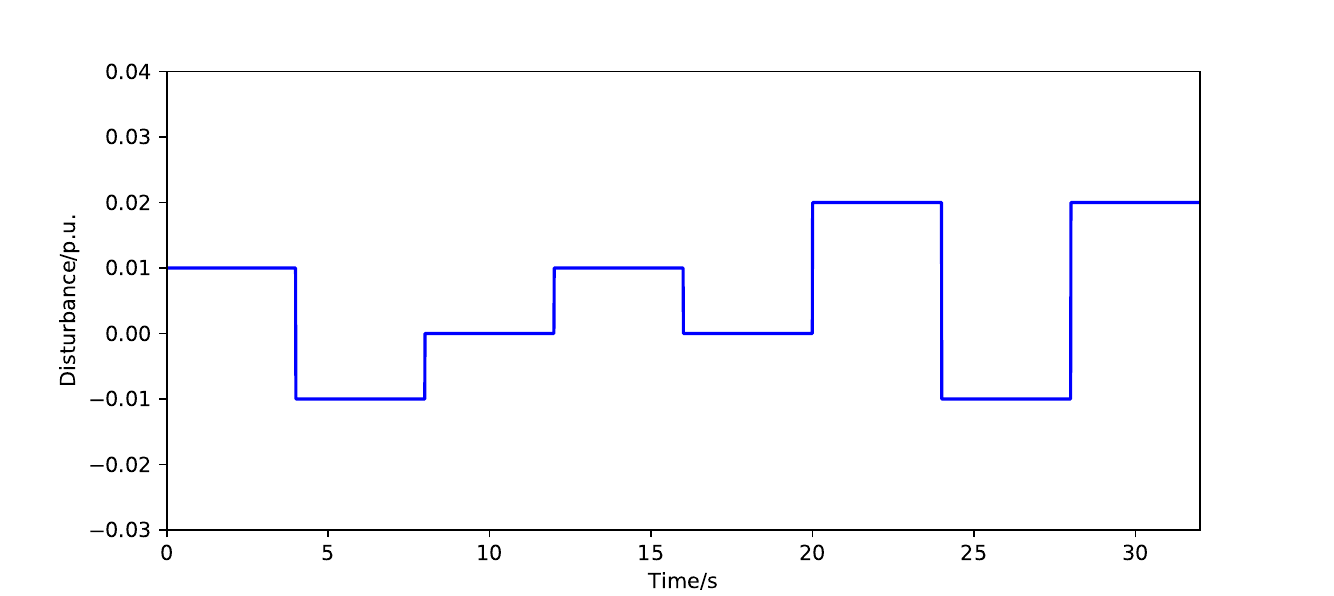}\\
  \caption{The step signal employed as load disturbance.}
  \label{figdisturbance}
\end{figure}

\begin{figure}[]
  \centering\includegraphics[width=9.5cm]{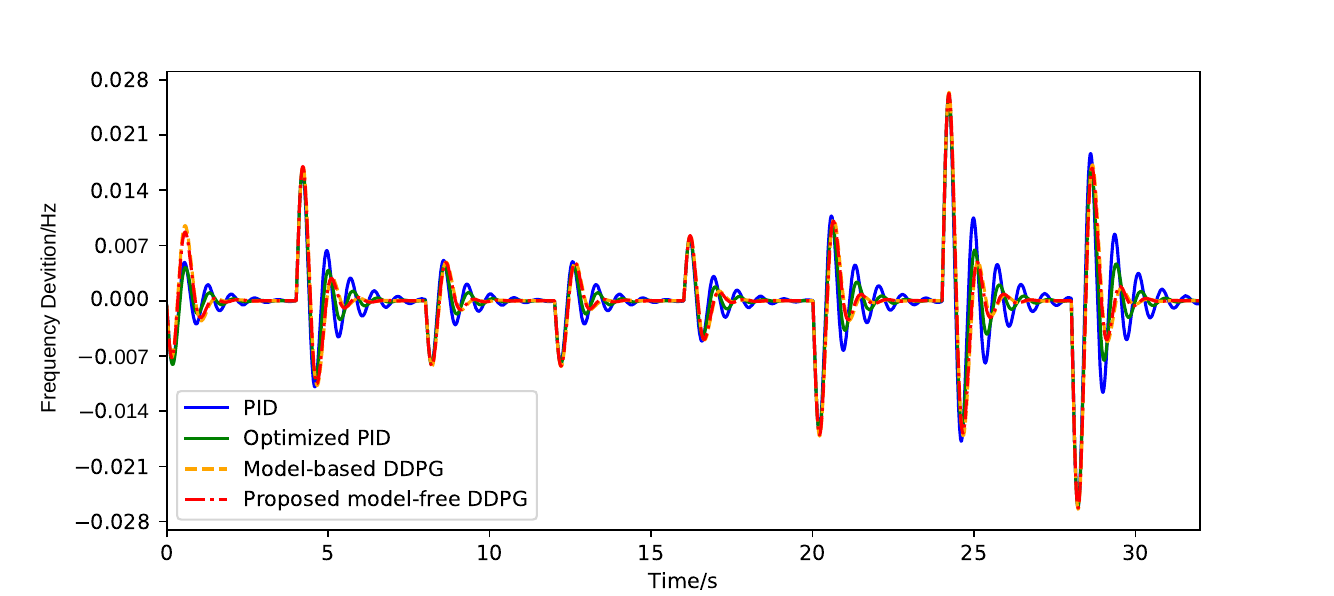}\\
  \caption{{Dynamic responses of frequency deviation on {linearized} power system with load disturbance.}}
  \label{fig4}
\end{figure}

\begin{figure}[]
  \centering\includegraphics[width=9.5cm]{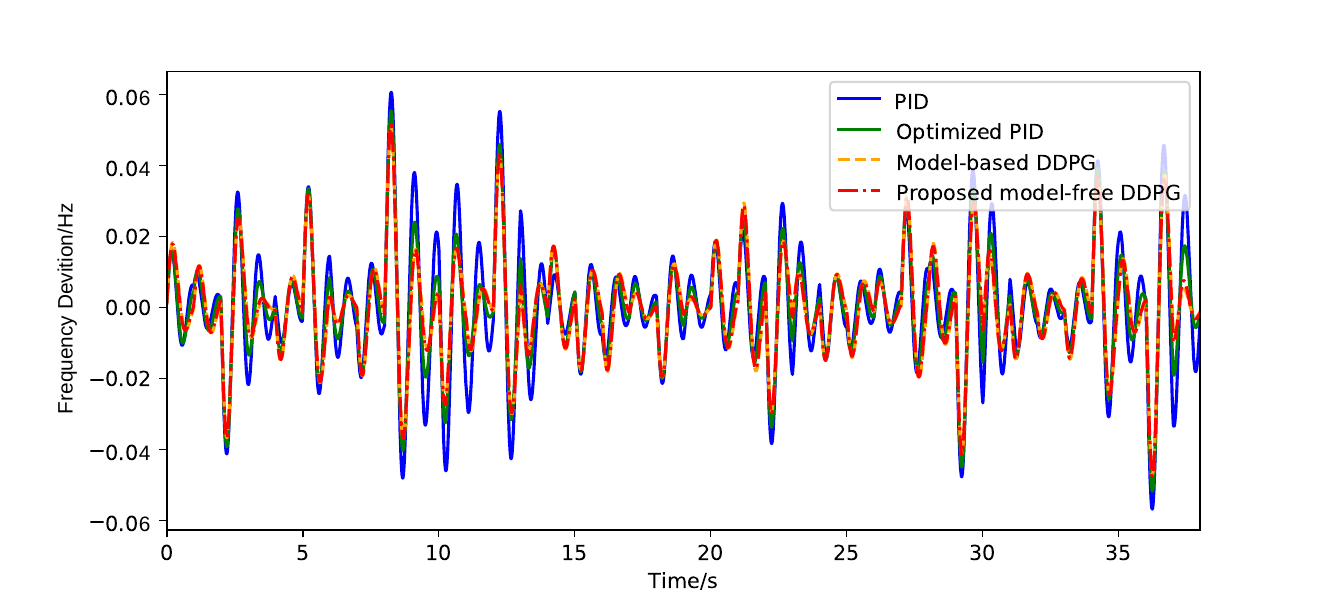}\\
  \caption{{Dynamic responses of frequency deviation on {linearized}  power system integrated with a wind farm.}}
  \label{fig5}
\end{figure}
 \begin{table}[]
  {
 \caption{Performance  {comparison} on linearized  power system.}
 \label{tab2}
 \renewcommand\arraystretch{2.5}
 \begin{tabular}{m{66pt}<{\centering}|m{31pt}<{\centering} m{48pt}<{\centering} m{55pt}<{\centering}}
   \hline
   \hline
    Method                                      &Action value $Q_\mu$   &Mean absolute of $\Delta f$    & Largest variation of $\Delta f$(Hz)  \\ \hline
    PID                                         &  -6.7520          &  0.0307               & 0.0505                            \\ \hline
    Optimized PID\cite{daneshfar2012multiobjective}& -4.7185        & 0.0251                & 0.0461                            \\ \hline
    Model-based DDPG\cite{yan2018data}          & -4.2725           & 0.0240                & 0.0423                            \\ \hline
    Proposed model-free DDPG                    & -4.2335           & 0.0237                & 0.0422                            \\\hline
   \hline
 \end{tabular}
 }
 \end{table}

Fig. \ref{otherfigure_linearreward} compares the training performance of the model-based and model-free DDPG algorithms on a linearized power system model. The orange line represents the model-based DDPG algorithm, while the red line corresponds to the model-free DDPG algorithm.
Both algorithms show an increasing trend in the performance metric $\mathcal{R}$ during training, indicating convergence towards improved control strategies beyond the initial  fine-tuned PID. Both algorithms achieve similar optimal control policy,  demonstrating comparable final training outcomes where the proposed model-free DDPG algorithm is slightly superior.

The step signal set as different values, as shown in Fig. \ref{figdisturbance},  is employed as load disturbance at different time points for the frequency response test of the linearized power system.
The $\Delta f$ curves of the closed-loop system with the  PID controller, the optimized PID controller, the model-based DDPG controller, and the proposed model-free DDPG controller are shown in Fig. \ref{fig4}. The $\Delta f$ curves of the closed-loop system integrated with a wind farm whose rated power is 0.02 p.u. for different controllers {are} compared in Fig. \ref{fig5}.  The figures {show} that compared with other controllers, $\Delta f$ fluctuations are reduced and $\Delta f$ deviations are smoothed significantly under the effect of the model-free DDPG controller. The numerical comparison result is shown in Table II, where the three quantified indices (action value $Q_\mu$, mean absolute of $\Delta f$, and largest variation of $\Delta f$) indicate the proposed method has noticeable improvement over the PID controller and the optimized PID controller. Table II also shows that the model-based DDPG method slightly performs the model-based one. Their overall performances are pretty close, proving the correctness of the trained emulator network and the calculated policy gradient.

\subsection{Results on nonlinear LFC Model}
To verify the advantages of the proposed model-free DDPG-based controller for nonlinear power systems, the performance comparison between some existing controllers and the proposed controller on the LFC model considering nonlinear generator behaviors (0.06\% GDB and 0.0017 p.u./s GRC) is presented in this subsection.

\begin{figure}[]
  \centering
  \includegraphics[width=8.9cm]{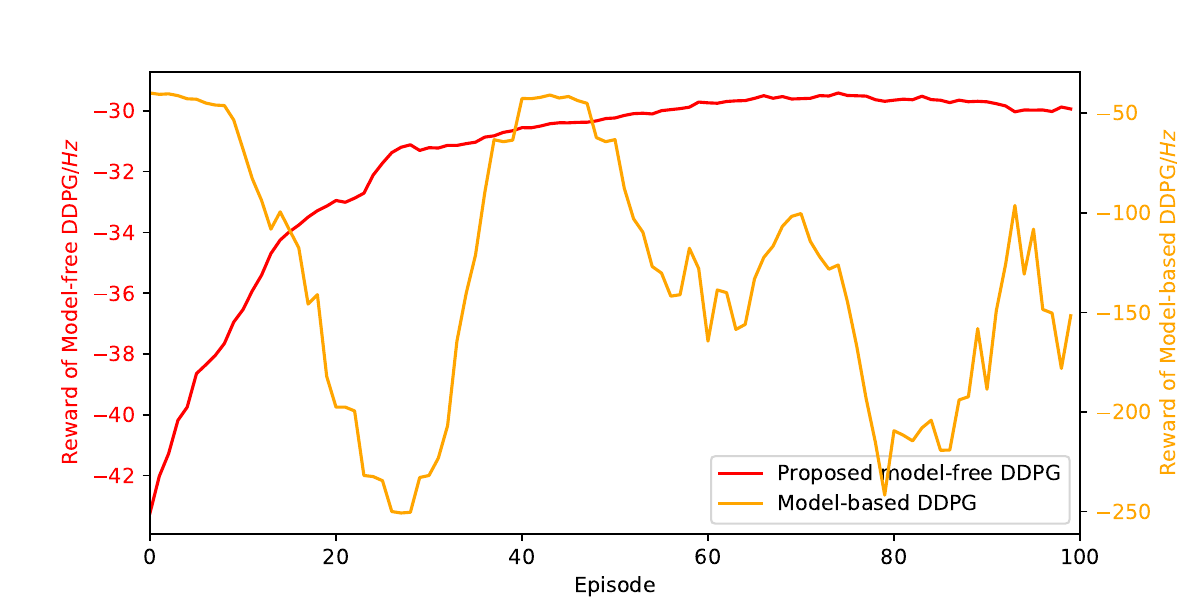}\\
  \caption{{The rewards during training process on nonlinear power system.}}
  \label{otherfigure_nonlinearreward}
\end{figure}

Fig. \ref{otherfigure_nonlinearreward} shows the training performance of the model-based DDPG (orange line) and model-free DDPG (red line) algorithm on a nonlinear power system model. As the number of training iterations increases, the model-based agent's performance, measured by $\mathcal{R}$, exhibits oscillations, suggesting a failure to converge to a superior control policy. This limitation renders the model-based DDPG algorithm unsuitable in scenarios where a complete understanding of the LFC model is not feasible.
In contrast, the model-free DDPG algorithm consistently improves its reward with an increasing number of training iterations, indicating the achievement of an optimized control policy.
In the model-free DDPG algorithm, the emulator network plays a critical role in guiding the update of the actor network with the help of the well-defined action-value function. Furthermore, the utilization of zeroth-order optimization addresses the challenge commonly encountered in DNN and significantly enhances the stability of the training process for the agent.
The model-free DDPG algorithm showcases remarkable adaptability and performs exceptionally well in overcoming the challenges posed by the nonlinear LFC model.

\begin{figure}[]
  \centering\includegraphics[width=9.5cm]{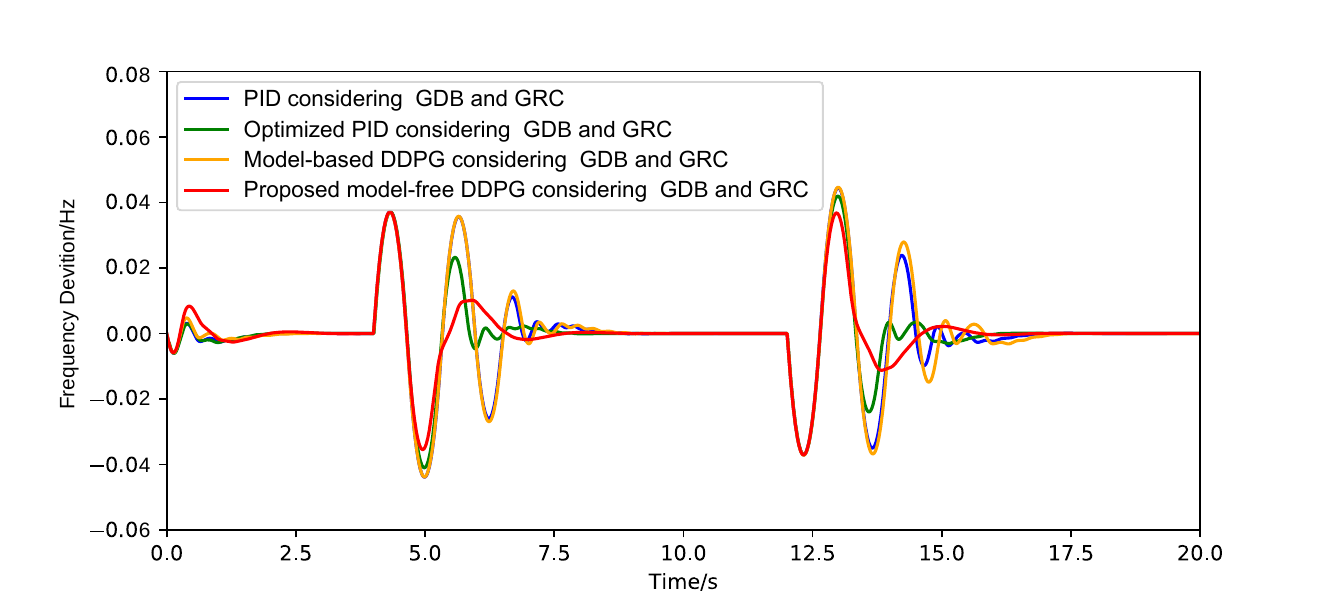}\\
  \caption{{Dynamic responses of frequency deviation on nonlinear power system with load disturbance.}}
  \label{fig6}
\end{figure}
\begin{figure}[]
  \centering\includegraphics[width=9.5cm]{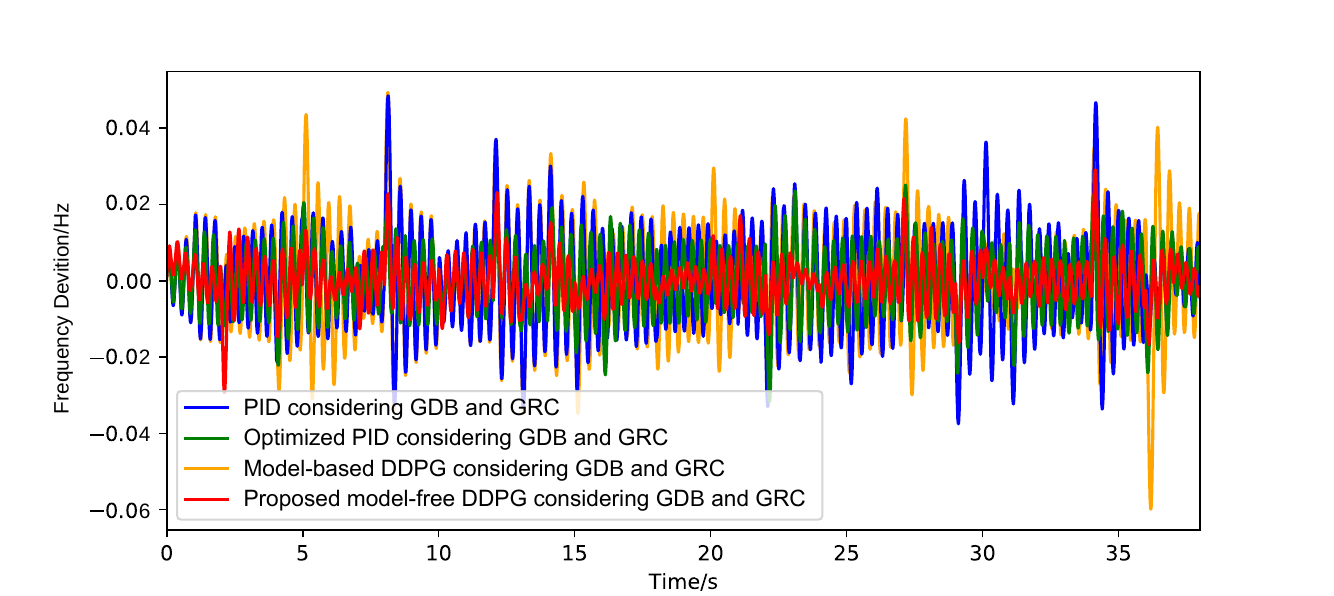}\\
  \caption{{Dynamic responses of frequency deviation on nonlinear power system integrated with a wind farm.}}
  \label{fig8}
\end{figure}
\begin{figure}[]
  \centering\includegraphics[width=9.5cm]{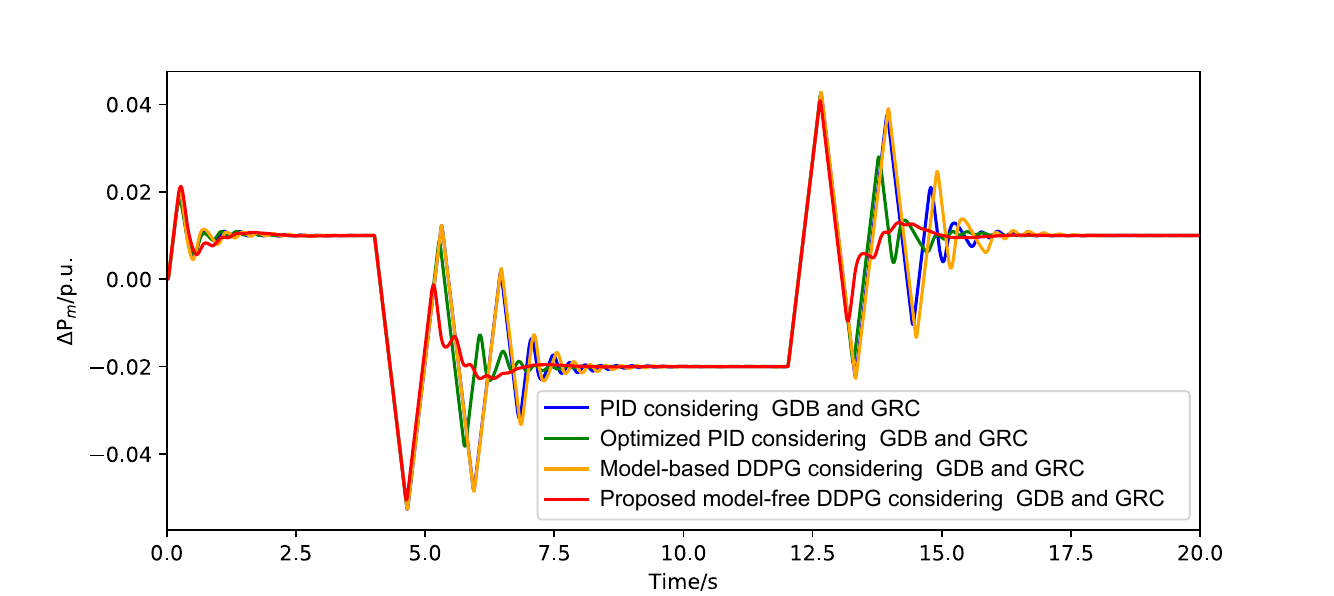}\\
  \caption{{Dynamic responses of generator active power outputs on nonlinear power system with load disturbance.}}
  \label{fig7}
\end{figure}
\begin{table}[]
  {
 \caption{Performance  {comparison} on nonlinear power system  considering
GDB and GRC.}
 \label{tab3}
 \renewcommand\arraystretch{2.5}
 \begin{tabular}{m{73pt}<{\centering}|m{31pt}<{\centering} m{48pt}<{\centering} m{55pt}<{\centering}}
   \hline
   \hline
    Method                                      &Action value $Q_\mu$   &Mean absolute of $\Delta f$    & Largest variation of $\Delta f$(Hz)  \\ \hline

    PID                         & -2.2671           & 0.0167                & 0.0370                            \\\hline
    Optimized PID \cite{daneshfar2012multiobjective}& -1.5662           & 0.0112                & 0.0349                            \\\hline
    Model-based DDPG \cite{yan2018data}& -2.3928           & 0.0117                & 0.0372                            \\\hline
    Proposed model-free DDPG    & -1.2369           & 0.0109                & 0.0309                            \\
   \hline
   \hline
 \end{tabular}
 }
\end{table}

Fig. \ref{fig6} shows  $\Delta f$ curves of the nonlinear closed-loop power system with a 0.03 p.u. step load decrease appearing at $t=4$s and a 0.03 p.u. increase appearing at $t=12$s for different LFC methods. Fig. \ref{fig8} shows  $\Delta f$ curves of the nonlinear closed-loop power system integrated with a wind farm whose rated power is 0.02 p.u. for different LFC methods. Fig. \ref{fig7} shows the active power output of the generator for different LFC methods considering GDB and GRC.
It is evident that in Fig. \ref{fig6},  Fig. \ref{fig8}, and Fig. \ref{fig7}, the $\Delta f$ curves and the active power output curve for the proposed model-free DDPG-based controller have better convergence effects than other $\Delta f$ curves, which proves the effectiveness and superiority of the proposed method for nonlinear power systems.

Meanwhile, Table \ref{tab3} presents a detailed quantitative numerical comparison. It is observed from Table \ref{tab3} that
the three quantified indices (action value $Q_\mu$, mean absolute value of $\Delta f$  and largest variation of $\Delta f$) of the proposed method are $-1.2369$, $0.0109$ and $0.0309$  respectively,
indicating that the proposed method has obvious improvement over the PID controller (ameliorated by 45.4\%, 34.7\%, and 16.4\%, respectively), the optimized PID controller (ameliorated by 21.0\%, 2.7\%, and 11.4\%, respectively), and the model-based DDPG controller (ameliorated by 48.3\%, 6.8\%, and 16.8\%, respectively).
Therefore, it is observed
that the proposed model-free DDPG method has obvious advantages over the model-based one under the nonlinear case.
The main reason is that the policy gradient calculated by the model-based DDPG method depends on the accuracy of the power system model, which may not capture the nonlinear characteristics of the generator and lead to a lack of representation of these features in the training process.  In contrast, the proposed method establishes the emulator network to extract nonlinear features and obtains appropriate policy gradient.


\section{Conclusion}\label{sec5}
This paper has presented a novel model-free LFC method based on the DDPG framework for nonlinear power systems.
A deep neural network named emulator network is established for power system emulation with a historical database. With the defined action-value function, the emulator network takes the place of the critic network to evaluate the control actions and guide the updating process of the actor network.
Zeroth-order optimization is applied to estimate the policy gradient, thereby avoiding the unstable updating of the agent caused by the gradient backpropagation process on multiple consecutive layers of deep neural network.
By utilizing the estimated policy gradient, the LFC agent is optimized to an optimal control policy status and achieves minimizing frequency deviations effectively.
Numerical simulations demonstrate the proposed method can generate more appropriate control commands and performs better than some existing LFC methods for both linearized and nonlinear power systems.
{
Following this work, we will further investigate the design of multi-agent deep reinforcement learning controller for multi-area load frequency control problems in the future. }

\bibliographystyle{ieeetr}
\bibliography{ref-LFC}
\end{document}